\documentclass[twocolumn,showpacs,preprintnumbers,amsmath,amssymb,floatfix]{revtex4}

\usepackage{graphicx}
\usepackage{dcolumn}
\usepackage{bm}

\begin{document}

\preprint{APS/123-QED}

\title{
Phase Diagrams and Charge-Spin Separation in\\ Two and Four Site
Hubbard Clusters}

\author{Armen~N.~Kocharian} 

\affiliation{%
Department of Physics and Astronomy, California State University,
Northridge, \\ CA 91330-8268
}%

\author{Gayanath~W.~Fernando}
\affiliation{Department of Physics, University of Connecticut,\\
 Storrs, CT 06269 and IFS, Hantana Rd., Kandy, Sri Lanka
}%

\author{James~W.~Davenport}
\affiliation{Computational Science Center, Brookhaven National Laboratory,\\
Upton, NY 11973}

\begin{abstract}
Charge spin separation, pseudogap formation and phase diagrams
are studied in two and four site Hubbard clusters using analytical
diagonalization and grand canonical ensemble method in a
multidimensional parameter space of temperature, magnetic field,
on-site Coulomb interaction ($U\ge 0$), and chemical potential. The
numerically evaluated, exact expressions for charge and spin
susceptibilities provide clear evidence for the existence of
 true gaps in the ground state and pseudogaps in a
limited range of temperature.
 In particular, Mott-Hubbard type charge crossover,
spin pseudogap and magnetic correlations with antiferromagnetic
(spin) pseudogap structure for two and four site clusters closely
resemble the pseudogap phenomena and the
normal-state
phase diagram in high T$_c$ superconductors.

\end{abstract}
\maketitle

Understanding the effects of electron correlations and 
pseudogap
phenomena~\cite{RVB,Nature,Timusk,Marshall,Kivelson_Review,Andrea}
in the cuprate superconductors comprising of many different phases
is regarded as one of the most challenging  problems in condensed
matter~\cite{Anderson}. Although the experimental determination of
various inhomogeneous phases in cuprates is still somewhat
controversial~\cite{Tallon}, the underdoped high T$_c$
superconductors are often characterized by crossover temperatures
below which excitation pseudogaps in the normal-state are seen to
develop~\cite{Zachar}. There is also compelling evidence for the
existence of  quantum critical points (QCPs) in
underdoped~\cite{Obertelli,Mang,Dagan} and optimally doped
materials~\cite{Boebinger} as observed in resistivity measurements
in Nd$_{2-x}$Ce$_x$CuO$_{4\pm\delta}$,
Pr$_{2-x}$Ce$_x$CuO$_{4-\delta}$ and La$_{2-x}$Sr$_x$CuO$_4$.

The charge-spin separation~\cite{Emery,Tsvelik}, clearly
identifiable Mott-Hubbard (MH), antiferromagnetic and spin
crossovers contain generic features which appear to be common for
small clusters and large thermodynamic systems~\cite{JMMM}.
Studies of conventional Mott-Hubbard and magnetic phase
transitions~\cite{Matt,Kimball,Sokol,Khom,Kotliar} have tended to
concentrate on macroscopic systems containing a large, and
essentially an infinite number of particles. In attempts to
address some of the above, the Hubbard model has been discussed
within the exact Lieb-Wu (LW) equations  in one dimension
(1d)~\cite{Lieb1,Tak2,Matt1} and a wide variety of 
approximation schemes in higher
dimensions~\cite{Jarrell,Jarrell1,Moukouri}. Most theories
originating from the Bethe-{\sl anstaz}, such as LW, involve
coupled nonlinear integral equations that have to be solved
numerically for every set of parameters. Numerical uncertainties
associated with such solutions severely limit their applications
when calculating subtle features at intermediate values of
temperature and other parameters. Although some properties of
Hubbard clusters have been
calculated~\cite{Shiba,Chen,Gros,Pastor,Joel,Avella,Schumann},
many questions remain with regard to
microscopic origins of charge-spin separation and
pseudogap behavior, short range correlations and weak
singularities ({\it crossovers}) at finite
temperature~\cite{Lieb3,Nagaoka}.

In this work, we study
the phase diagrams for the
two and four site Hubbard clusters~\cite{JMMM} using  analytical
diagonalization combined with the grand canonical ensemble in a
multidimensional parameter space of temperature $T$, magnetic
field $h$, on-site Coulomb interaction $U\ge 0$, and chemical
potential $\mu$ (with  hopping parameter $t=1$). Our calculations
for finite clusters are based on the exact analytical expressions
for the eigenvalue $E_{n}$ of the $n^{th}$ many-body eigenstate, grand
partition function Z (where the number of particles $N$ and the
projection of spin $s^z$ are allowed to fluctuate) and its
derivatives without taking the thermodynamic limit and hence
involve no approximations. The grand canonical potential
$\Omega_U$ for interacting electrons is
\begin{eqnarray}
\Omega_U=-{T}\ln \!\sum\limits_{n\leq N_H} e^{{-\frac {E_{n}-\mu
N_{n} - hs^{z}_n} T} }, \label{OmegaU}
\end{eqnarray}
where $N_n$ and $s^{z}_n$ are the number of particles and the
projection of spin in the $n^{th}$ state respectively. The
dimension $N_H$ of the Hilbert space in (\ref{OmegaU}) depends on
the number of sites, satisfying $N_H=4^2$ for the two site
cluster, and $N_H=4^4$ for the four site cluster. The electron
charge susceptibility ${\chi_c}(\mu)$ or the corresponding
thermodynamic density of states,  $\rho(\mu)={{\frac {\partial
{\left\langle {N(\mu)}\right\rangle}} {\partial \mu}}}$, describes
the local spectral characteristics of charge excitations and
fluctuations in the number of electrons
\begin{eqnarray}
\left\langle N^2\right\rangle-\left\langle
N\right\rangle^2=T{{\frac {\partial {\left\langle
{N(\mu)}\right\rangle}} {\partial \mu}}}.
\label{fluctuation_number}
\end{eqnarray}

The spin susceptibility $\chi_s(\mu)$ or spin density of states
$\sigma(\mu)={{\frac {\partial \left\langle s^{z}\right\rangle}
{\partial h}}}$ describes the local spin excitations as parameters
$h$, $\mu$ or $T$ are varied, where spin fluctuations
$\left\langle ({\Delta s^{z}})^2\right\rangle$ closely follow the
variation of spin susceptibility $\chi$ with respect to $\mu$ and
$h$
\begin{eqnarray}
\left\langle {(s^{z})}^2\right\rangle-\left\langle
s^{z}\right\rangle^2=T{{\frac {\partial \left\langle
s^{z}\right\rangle} {\partial h}}}. \label{fluctuation_mag}
\end{eqnarray}

\begin{figure} 
\begin{center}
\includegraphics*[width=15pc]{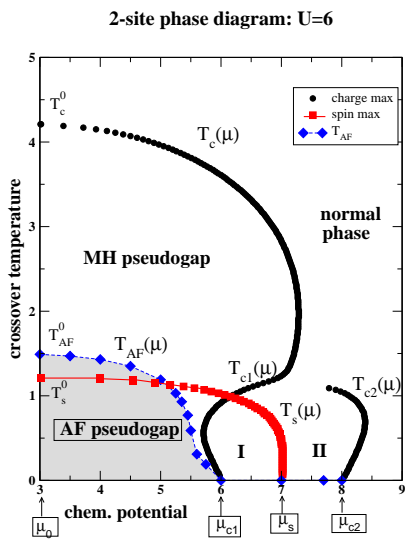}
\end{center}
\caption {Temperature $T$ vs chemical potential $\mu$ phase
diagram for the two site cluster at $U=6, h=0$. The
antiferromagnetic pseudogap phase is shaded and bounded above by
$T_{AF}(\mu)$. It vanishes in regions I and II, which are charge
bifurcation phases  representing (pseudo)  charge gaps at $T>0$.
The charge and spin crossover temperatures $T_c(\mu)$ and $T_s
(\mu)$ are obtained from maxima in charge and spin
susceptibilities. Since there is particle-hole symmetry, only
$\mu$ values above half-filling, i.e. $\mu > \mu_0$,  are shown. Note
that in regions I and II, there is strong charge-spin separation.}
\label{fig:phase-diagram-2}
\end{figure}
\begin{figure} 
\begin{center}
\includegraphics*[width=15pc]{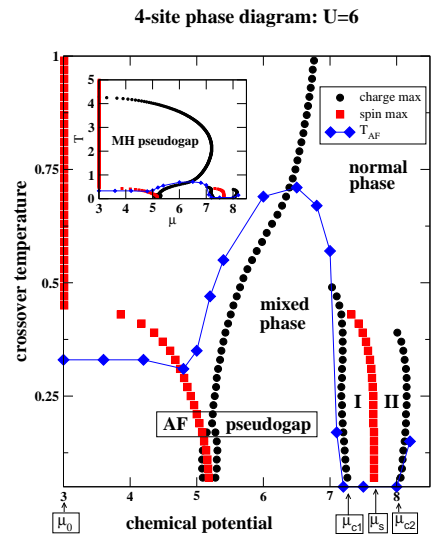}
\end{center}
\caption {Temperature $T$ vs chemical potential $\mu$ phase diagram
 for the four site cluster at $U=6, h=0$. Regions I and II are quite similar to the ones found in the two
site cluster, again showing strong charge-spin separation.
However, note the (new) mixed phase which consists of charge, spin
and AF pseudogaps, that is present in this cluster. The inset
shows the same diagram with a larger scale for the crossover
temperature. Labels of crossover temperatures are suppressed for
clarity.}
\label{fig:phase-diagram-4}
\end{figure}

Once all the many-body eigenvalues of the Hubbard clusters are
known, it is straightforward to calculate the ground state
properties and these results are reported elsewhere~\cite{JMMM}.
Using the same eigenvalues, we have evaluated the {\sl exact}
grand partition function and thermal averages such as
magnetization and susceptibilities numerically as a function of
the set of parameters $\{T,h,\mu,U\}$. Using maxima and minima in
spin and charge susceptibilities, phase diagrams in a $T$  vs
$\mu$ plane for any $U$ and $h$ can be 
constructed.

Among many interesting results rich in variety for $U>0$, sharp transitions
are found between phases with true charge and spin gaps in the
ground state; for infinitesimal $T>0$, these gaps are transformed
into `pseudogaps' with some nonzero weight between peaks (or
maxima) in susceptibilities monitored as a function of doping
(i.e. $\mu$) as well as $h$. We have also verified the well known
fact that the low temperature behavior in the vicinity of
half-filling, with charge and spin pseudogap phases coexisting,
represents an antiferromagnetic insulator in the Hubbard
clusters~\cite{JMMM}. However, {\sl away from half filling}, we
find very intriguing behavior in thermodynamical charge and spin
degrees of freedom.
At low temperature, new peak structures in the charge
$\chi_c(\mu)$ and zero magnetic field spin $\chi_s(\mu)$
susceptibilities are observed to develop~\cite{JMMM}; between two
consecutive peaks, there exists a pseudogap in charge or spin
degrees. Opening of such distinct and separated pseudogap regions
at low temperature, the signatures of corresponding charge and
spin separation, is seen in both the two and four site clusters,
away from half filling.

In Fig.~\ref{fig:phase-diagram-2}, we show the phase diagram for
the two site cluster at $U=6$ and $h=0$, where $T_c(\mu)$ describes the high
temperature, MH insulator-metal transition as $\mu$ varies. At low
temperature, this dependence bifurcates and  $T_{c1}(\mu)$ and
$T_{c2}(\mu)$ denote the crossover temperatures at lower and
higher doping (respectively) corresponding to similar MH like
transitions. The low temperature regions I and II, bounded by
these crossover temperatures, define a charge bifurcation phase
which persists  up to a high doping level, $\mu_{c2}$. In this
charge pesudogap, there are strong spin fluctuations centered
around $\mu_s$ at low temperature which decay when either $\mu\to
\mu_{c1}$ or $\mu\to \mu_{c2}$ (i.e. near the boundaries where
charge gaps melt). In the same figure, the antiferromagnetic
crossover temperature $T_{AF}(\mu)$ is obtained by monitoring the
spin susceptibility peaks as a function of the applied field at a
given chemical potential $\mu$,
as seen in measurements of N{\'e}el temperature doping
dependence~\cite{Mang}.
If $h_c$ denotes the field value
corresponding to the peak closest to zero, $T_{AF}$ is defined as
the temperature at which the critical field $h_c(\mu)$ for the
onset of magnetization vanishes (melting the AF spin gap).
$T_{AF}$ can also obtained from zero field staggered spin
susceptibility.
The spin crossover temperature ${T_s}(\mu)$, associated with the
opening of the zero-magnetic-field spin pseudogap in
Fig.~\ref{fig:phase-diagram-2}, denotes the temperature below
which  strong spin pseudogap correlations are observed to develop.
This has been suggested as a precursor to
superconductivity~\cite{Shen}. Note that spin fluctuations are
significantly suppressed below ${T_s}$ and how these crossover
temperatures and critical $\mu$ values define various regions such
as, a) charge pseudogap, b) AF pseudogap, c) spin pseudogap and d)
normal phases.

The four site phase diagram, Fig.~\ref{fig:phase-diagram-4} with $U=6$ and $h=0$,
appears to be more complex with several charge bifurcation
phases above (or below) half filling, starting at the electron (hole)
doping level near $1/8$ filling. However, there is a clear
self-similarity in the right-most bifurcation region for the
2-site and the 4-site clusters  (Figs.~\ref{fig:phase-diagram-2}
and ~\ref{fig:phase-diagram-4}). This bifurcation region retains
strong charge pseudogap stability as in the two site cluster.
 We believe that these regions of phase space, with strong spin and charge fluctuations, are quite
relevant to the high T$_c$ cuprates and other similar materials,
where doping of electrons or holes introduces dramatic changes in
their physical properties.

We have followed the behavior of pseudogaps in this region as a
function of the on-site Coulomb repulsion $U$. With increasing
$U$, several features are observed to develop; (a) separation of
charge and spin boundaries
%
away from half-filling, (b) opening of a pseudo charge gap,
(c) large spin fluctuations inside this charge gap
region.
 This phase with a
charge gap closely resembles the inhomogeneous commensurate phase
found in Ref.\cite{Zachar}, where the mobile holes in the
quasi-one dimensional structures (stripes) acquire a spin gap in
spatially confined Mott-insulating regions~\cite{Zachar,Nature}.

The doping dependence of the normal-state {\sl spin pseudogap}
shows that at low temperature, it is stable in the underdoped
regime, persists at optimal doping and disappears in the overdoped
regime at $\mu = \mu_s$. In the phase diagrams in
Figs.~\ref{fig:phase-diagram-2} and ~\ref{fig:phase-diagram-4} at
zero temperature, we notice several quantum phases and
corresponding quantum critical points (QCPs), at $\mu=\mu_{c1},
\mu_s, \mu_{c2}$. A comparison of the two phase diagrams for the
two and four site clusters, in Figs.~\ref{fig:phase-diagram-2} and
~\ref{fig:phase-diagram-4}, reveals many common features in
addition to the QCPs. As $\mu$ increases, there is a sharp
transition in both clusters at $\mu_{c1}$ from a Mott-Hubbard
antiferromagnetic insulator into a phase with charge and spin
separation and gaps. A similar behavior has been observed
experimentally ~\cite{Mang,Dagan} in the cuprates. 
At critical doping $\mu_{c1}$, the antiferromagnetic phase
disappears and at higher doping, in regions I and II, the spin and
charge pseudogap phases coexist with one another independent of
how strong $U$ is. These two regions are separated by a boundary
where the spin gap vanishes. At critical doping $\mu_s$ with T$_s
\to 0$, the zero spin susceptibility $\chi_s$ at zero temperature
exhibits a sharp maximum. Thus the behavior of the critical
temperature $T_s (\mu)$, which falls abruptly to zero at critical
doping $\mu_s$,
implies~\cite{Obertelli,Tallon}
that the pseudogap can exist independently of possible  superconducting
pairing. As mentioned above, as $T\to 0$ and $\mu = \mu_s$, the
spin gap disappears while the charge gap prevails up to $\mu = \mu_{c2}$.
 Up on further doping, the charge gap vanishes at
$\mu_{c2}$ and Fermi liquid behavior is restored due to full
charge-spin reconciliation. Figs.~\ref{fig:phase-diagram-2} and
~\ref{fig:phase-diagram-4} are consistent with the existence of
pseudogap phases
and quantum phase transitions at $\mu_s$
in the high $T_c$ superconductors,
 when the ground state spin gap disappears
~\cite{Marshall,Obertelli,Castellani,Andrea}. The QCPs separate
the spin pseudogap phase from the quantum spin liquid state,
coexisting with the charge pseudogap. We have also seen that a
reasonably strong magnetic field has a dramatic effect (mainly) on
the QCP at $\mu_s$.

The spin susceptibility $\chi_s$ in the underdoped pseudogap
region I, defined by $\mu_{c1}\leq \mu \leq \mu_s$, displays an
anomaly at temperature ${T_s}$ below which the spin degrees of
freedom become suppressed. The gapped spin excitations imply that
spin singlet states which exist between ${T_s}$ and $T_{c1}$ in
Fig.~\ref{fig:phase-diagram-2} can be considered as pre-formed
pairs with properties different from FL. As temperature decreases,
approaching  $T_{c1}$ from above, $\chi_s$ decreases while
$\chi_c$ increases due to strong charge fluctuations in the
vicinity of $T_{c1}$ signaled by a sharp peak in the excitation
spectrum, consistent with ARPES measurements~\cite{Ding1}. The
thermal fluctuations close to the boundaries $T_{c1}$ of the
charge bifurcation phase destroy the charge pseudogap  and provide
conditions at which the spin pseudogap state coexists with a
charge liquid background.
\begin{figure} 
\begin{center}
\includegraphics*[width=10pc]{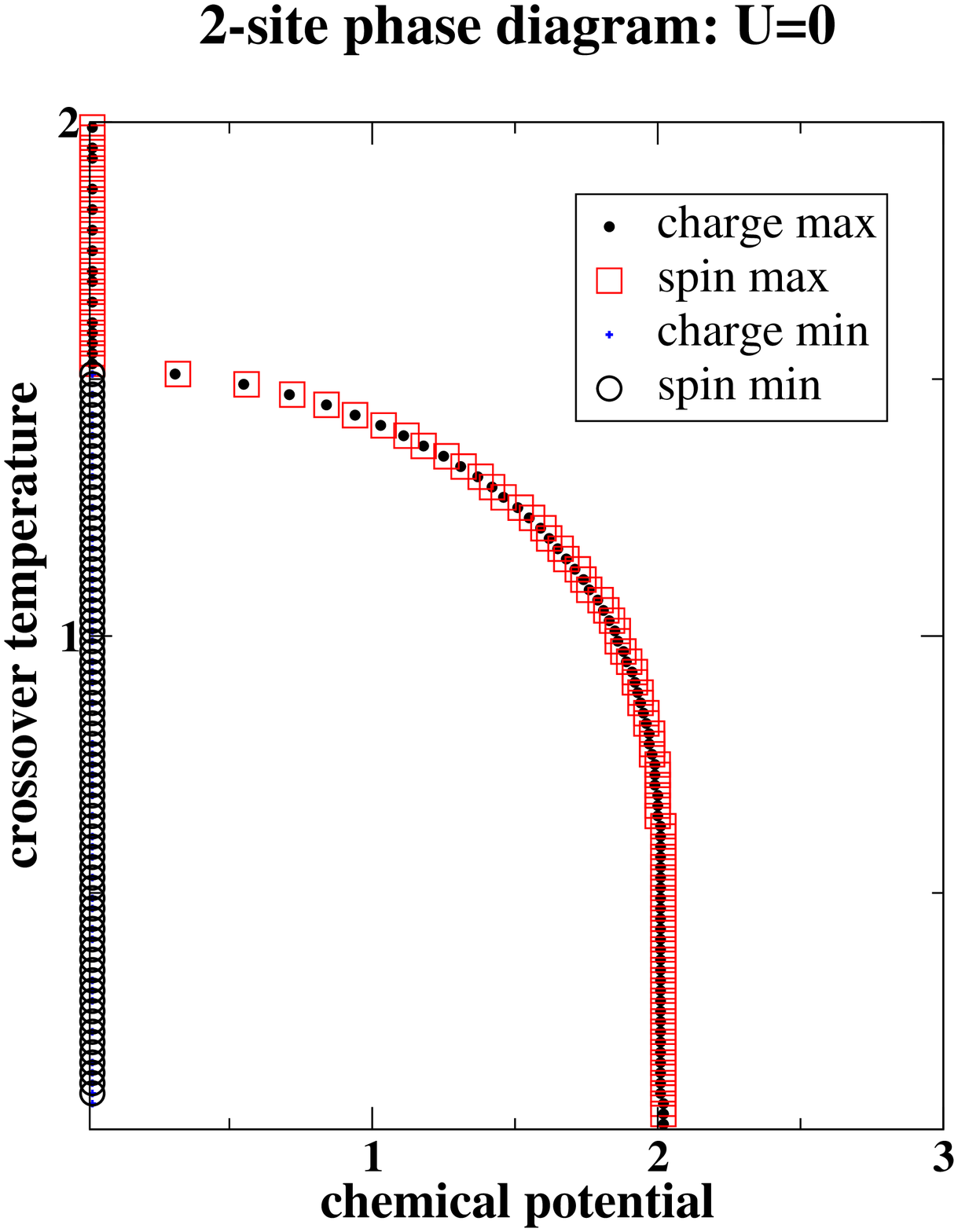}
\includegraphics*[width=10pc]{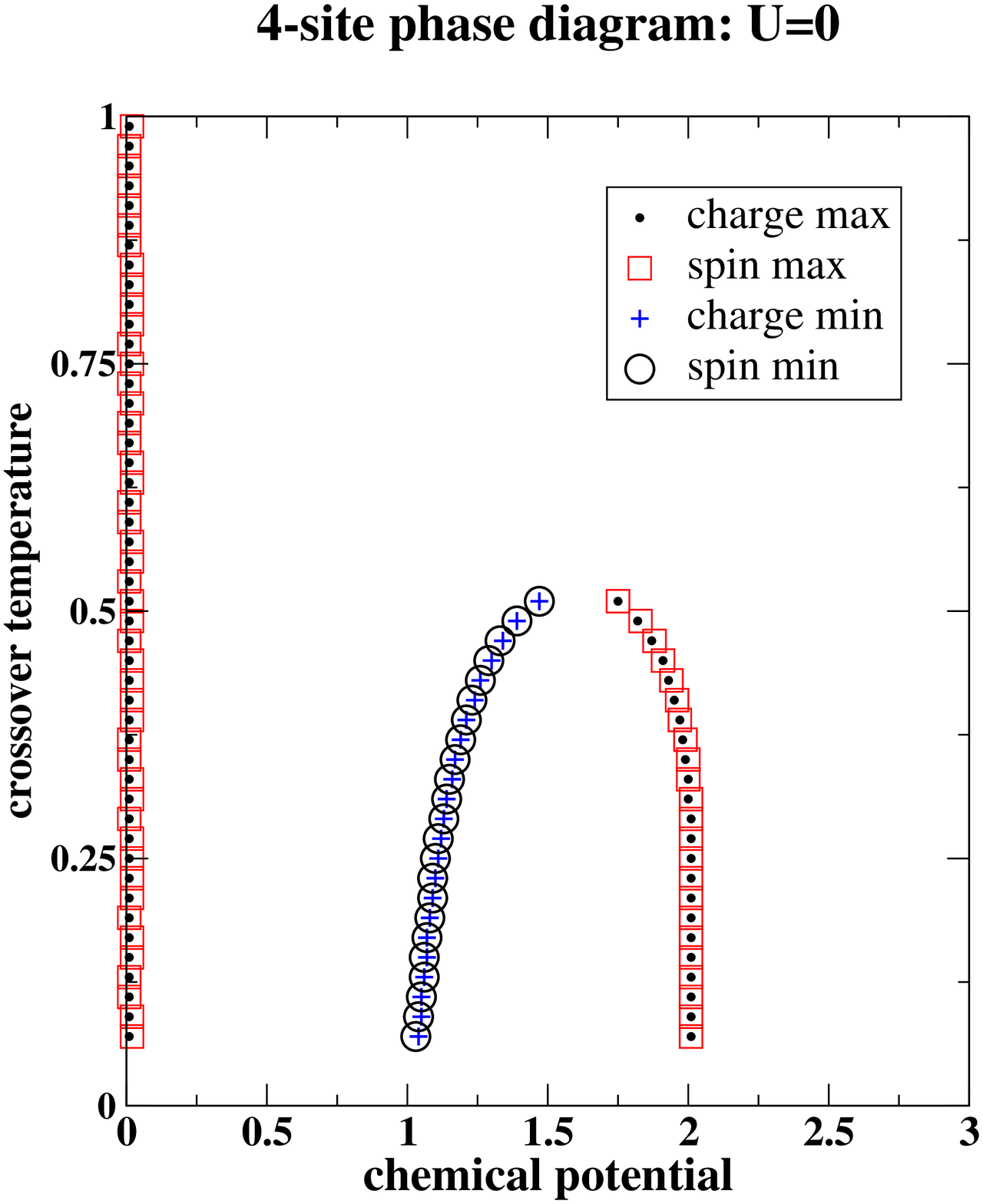}\hfill
\end{center}
\caption {The single particle or `noninteracting' ($U=0$) case,
illustrating the charge and spin, peaks and valleys for the two
and four site clusters. Note how the charge and spin maxima and minima
in dos follow one another indicating the absence of charge-spin separation.}
\label{fig:phase_u0}
\end{figure}

In Figs.~\ref{fig:phase-diagram-2} and ~\ref{fig:phase-diagram-4},
we also show the crossover temperature $T_{AF}(\mu)$ deduced from
maxima in the antiferromagnetic spin fluctuations $\sigma(\mu)$.
The phase diagram for the two site cluster shows an island of
stability for the antiferromagnetic phase and the corresponding
 crossover temperature, $T_{AF}(\mu)$,  decreases monotonically as the
chemical potential $\mu$ increases. In the underdoped regime close
to half filling, the antiferromagnetic phase in the two site
cluster is fully separated from the charge bifurcation region. In
contrast, the behavior of $T_{AF}$ with increasing $\mu$ is
nonmonotonic for the four site cluster and there is  some overlap
(near $T_{AF}\approx T_{c1}$) in the underdoped regime between the
antiferromagnetic phase and a charge bifurcation  region. At
higher doping within the (right-most) bifurcation regime (i.e.
regions I and II), $T_{AF}\equiv 0$ and the charge-spin separation
is very similar to that seen in the two site cluster.

As an important footnote, we show how the charge and spin peaks
(and valleys) follow one another for both the two and four site
clusters when $U=0$ in Fig.~\ref{fig:phase_u0} (in sharp contrast
to the $U=6$ cases, in regions I and II where charge (as well as
spin) maxima and minima are well separated) indicating that there
is no charge-spin separation here. In this limiting, single
particle case for the two site cluster, there are no  temperature
driven charge bifurcation phases away from half filling and hence
the corresponding $T_{c1}\equiv 0$. In the entire range $-2t\leq
\mu\leq 2t$, the charge and spin fluctuations directly follow one
another, i.e. $T_{c1}=T_{c2}=T_s$ without charge-spin separation.

In summary, we have illustrated the phase diagram and the presence
of temperature driven crossovers, quantum phase transitions and
charge-spin separation for any $U\neq 0$ in the two and four site
Hubbard clusters as doping (or chemical potential) is varied.
Our bottom-up  approach and exact thermodynamics
for small clusters, when monitored as a function of doping,
displays the presence of clearly identifiable, temperature driven
crossovers into new phases and distinct transitions at
corresponding QCPs in the ground state, seen in large
thermodynamic systems.
It appears that the
short-range correlations alone are sufficient for pseudogaps to
form in small and large clusters, which
can be linked to the generic features of phase
diagrams in temperature and doping effects seen in the high $T_c$
cuprates. The pseudogap features and the variations of $\chi_c$,
$\chi_s$, $T_{AF}$ with $\mu$ as well as the existence of QCPs
suggest that the normal state spin singlet pseudogap, closely
linked to short range correlations, can also exist in small
clusters. The exact cluster solution shows how charge and spin
gaps are formed at the microscopic level and their behavior as a
 function of doping (i.e.
chemical potential), magnetic field and temperature. The pseudogap
formation can also be associated with the condensation of spin and
charge degrees of freedom below (spin and charge) crossover
temperatures $T_s$ and $T_{c1}$ (respectively). Finally, the two
and four sites clusters share very important intrinsic
characteristics with the high $T_c$ superconductors apparently
because in all these `bad' metallic high $T_c$ materials, local
interactions play a key role.

This research was supported in part by the U.S. Department
of Energy under Contract No. DE-AC02-98CH10886.

\end{document}